\documentstyle[12pt]{article}

\begin{document}

\author{Jos\'e I. Usera \\
Universidad Complutense de Madrid}
\title{A study of the consistency between noncommutative quantum mechanics and
Galilean isotropy.}
\date{1, 18, 2000}
\maketitle

\begin{abstract}
A demonstration is given that the simplest model of quantum mechanics
formulated on a plane non-commutative geometry endowed with a Galilean
symmetry group in which the position and linear momentum-variable
commutators are first order in the dynamical variables (and thus constitute
a true Lie algebra) is incompatible with the hypothesis of spacial isotropy.
\end{abstract}

{\it ``Civilization advances by extending the number of important operations
which we can perform without thinking about them.''}

Alfred North Whitehead.

\section{Introduction}

The proposal to generalise Heisenberg commutation relations in order to
include mutually noncommuting position variables\ is not a new avenue at
all. But in recent years, the argument has received a renewed momentum in
connection with the search for a valid generalization of quantum field
theory. It is well-known that this theory gives rise to infinite quantities
for whose removal we need to put to practice renormalisation procedures. One
of these, known as the Pauli-Villars dimensional regularisation, consists in
decreeing a cutoff in the range of values for linear momentum contributing
to the cross section. Thus, all ultraviolet contributions are removed, the
finite parte is extracted, and only at the end one makes the mentioned
cutoff in the momentum space go to zero. Such techniques are not totally
free from ambiguity and quantum field theory suffers from some problems at
least of a methodologic nature as is, e. g., the problem of the vacuum
energy (related to the value of the cosmologic constant in Einstein's
equations) or the non-renormalizability of the quantum theory of gravitation.

Among the alternative ideas that have been coming through, (as is the one of
assuming a string structure beyond Plank's length scale) ranks the theory of
non-commutative geometries, which occupies my attention in this work.

The fundamental idea of this theory, expressed in a way rather more prosaic
than usual in the specialised literature, stems from the generalisation of
Heisenberg commutation relations

\begin{equation}  \label{heisenberg}
\begin{array}{cc}
\left[ X_i,P_j\right] =i\hbar \delta _{ij}, & \left[ X_i,X_j\right] =\left[
P_i,P_j\right] =0
\end{array}
\end{equation}

to a more general one of the form

\begin{equation}  \label{noconmuta}
\begin{array}{cc}
\left[ X_i,X_j\right] \neq 0, & \left[ P_i,P_j\right] \neq 0
\end{array}
\end{equation}

The most famous consequence of (\ref{heisenberg}) is the impossibility to
prepare states such that, once the pertinent measurements have been
performed, lead to dispersion relations for $X_i$ and $P_j$ {\it %
simultaneously better than} those given by

\begin{equation}  \label{incertidumbreXP}
\Delta _\psi X_i\Delta _\psi P_j\geq \frac 12\hbar \delta _{ij}
\end{equation}

as results from the more general version for two noncommuting quantum
operators $A$ and $B$:

\begin{equation}  \label{incertidumbre2}
\Delta _\psi A\Delta _\psi B\geq \frac 12\left| \left\langle \psi \left| 
\left[ A,B\right] \right| \psi \right\rangle \right|
\end{equation}

This more general formulation, allows to generalise the uncertainty
principle in such a way that it precludes the possibility to construct
states with the position defined in a point-like manner. The fact that,
following (\ref{incertidumbre2}) for the instance $A=X_{i}$ and $B=X_{j}$
with $i\neq j$, it is impossible to define both coordinates with unlimited
precision, leads to ruling out all states having a localisation probability
density point-like arranged. Now, it has been pointed out in numerous
occassions the possibility that the infinities that plague quantum theories
come from an excessive idealisation consisting in the handling of such
``point-like particle states''. Should the hope for such a spacially
extended model be realised, the doors may be open to the existence of a
theory finite to every order (or, at least, renormalisable,) of which the
present point-particle quantum field theory, would be but a low-energy
version or an effective field theory.

\section{Noncommutativity and isotropy}

In the present work, I will investigate the consistency of a certain model
of non-commutative structure for the dynamical variables in a Galilean,
first-quantised quantum mechanical scenario. More in particular, I will
assume the commutation relations in such context are of the form

\begin{equation}  \label{c1}
\left[ X_i,P_j\right] =i\delta _{ij}+iu_{ijk}X_k+iv_{ijk}P_k
\end{equation}

\begin{equation}  \label{c2}
\left[ X_i,X_j\right] =i\alpha _{ij}+if_{ijk}X_k+ig_{ijk}P_k
\end{equation}

\begin{equation}  \label{c3}
\left[ P_i,P_j\right] =i\beta _{ij}+il_{ijk}X_k+im_{ijk}P_k
\end{equation}
where the numbers $\alpha _{ij}$ and $\beta _{ij}$ constitute real
skew-symmetric matrices and the structure constants $f_{ijk}$, $g_{ijk}$, $%
l_{ijk}$ and $m_{ijk}$ are also real and skew-symmetric in their first 2
indices. That the dynamical-variable commutators (operators) should lead
just to first-order terms in these is not, to be sure, a logical necessity
(they could be, e.g., polynomials in ${\bf X}$ and ${\bf P}$, or perhaps
more general functions). However, it seems the most natural hypothesis to
demand a closure condition or, in other words, for these dynamical variables
to form a proper Lie algebra. In order for this to be complied with, it is
necessary that such commutations produce linear combinations of the
operators we started from. As a matter of fact, the most general theory of
noncommutative geometries developed by Connes {\it et al.}, either assumes a
more general structure or does not make an issue of it at all.\ But notice
that the possibility of the occurrence of, e.g., quadratic terms in ${\bf X}$
or ${\bf P}$ would be far more serious a problem than it seems at first
sight. Indeed, the fact that such operators are not bounded in the
test-function space, renders every successive multiplication of any of them
problematic, as it forces us to expect very drastic changes in the domain.
Thus, the sheer writing of (\ref{heisenberg}) already meets problems in
ordinary quantum mechanics, for the respective domains of the identity
operator on the one hand and the mutually isomorfic ones ${\bf X}$ and ${\bf %
P}$ do not coincide. That is why, the rigorous way to write (\ref{heisenberg}%
) is actually

\begin{equation}  \label{heisenberg-Jammer}
\left[ X_i,P_j\right] \subseteq i\hbar \delta _{ij}
\end{equation}
in the understanding that every assignation from domain to image in the
l.h.s. are included among the assignations given by the r.h.s., but the
converse not being necessarily true. The successive multiplications of
position and linear momentum variables would then gradually ``erode'' the
domain in such a way that the operators corresponding to the dynamical
functions $f({\bf X},{\bf P})$ would not be well-defined anymore. Weyl's
version of (\ref{heisenberg-Jammer}) is another way of saying the same:

\begin{equation}  \label{Weyl}
\exp (i{\bf a}\cdot {\bf X})\exp (i{\bf b}\cdot {\bf P})=e^{-i{\bf a}\cdot 
{\bf b}}\exp (i{\bf b}\cdot {\bf P})\exp (i{\bf a}\cdot {\bf X})
\end{equation}
as results from being expressed in terms of operators free from
domain-related problems.

The central idea of the present work consists in exploring the successive
commutations of these operators\ under restrictions (\ref{c1})-(\ref{c3})
using the Leibniz rule for the commutator: 
\begin{equation}  \label{leibniz}
\left[ A,BC\right] =\left[ A,B\right] C+B\left[ A,C\right]
\end{equation}
that is: ``the commutator is a derivative''. We have then, on the one hand,
by directly using (\ref{c1})-(\ref{c3}):

\[
\left[ X_i,\left[ X_j,P_k\right] \right] =iu_{jkl}\left[ X_i,X_l\right]
+iv_{jkl}\left[ X_i,P_l\right] = 
\]
\[
-(u_{jkl}\alpha _{il}+v_{jki})-\left( u_{jkl}f_{ilm}+v_{jkl}u_{ilm}\right)
X_m-\left( u_{jkl}g_{ilm}+v_{jkl}v_{ilm}\right) P_m 
\]
and on the other hand, using (\ref{leibniz}), and also (\ref{c1})-(\ref{c3})
again

\[
\left[ X_i,\left[ X_j,P_k\right] \right] =\left[ X_i,X_jP_k\right] -\left[
X_i,P_kX_j\right] = 
\]

\[
\left[ X_i,X_j\right] P_k+X_j\left[ X_i,P_k\right] -\left[ X_i,P_k\right]
X_j-P_k\left[ X_i,X_j\right] = 
\]

\[
if_{ijl}\left[ X_l,P_k\right] +ig_{ijl}\left[ P_l,P_k\right] +iu_{ikl}\left[
X_j,X_l\right] +iv_{ikl}\left[ X_j,P_l\right] = 
\]

\[
-f_{ijk}-v_{ikj}-g_{ijl}\beta _{lk}-u_{ikl}\alpha
_{jl}-(f_{ijl}u_{lkm}+g_{ijl}l_{lkm}+u_{ikl}f_{jlm}+v_{ikl}u_{jlm})X_m 
\]

\[
-(f_{ijl}v_{lkm}+g_{ijl}m_{lkm}+u_{ikl}g_{jlm}+v_{ikl}v_{jlm})P_m 
\]
which, after identification of similar terms with respect to the former
expression ($I$, $X_i$ amd $P_i\ $should constitute a basis of our algebra),
allow us to obtain: 
\begin{equation}  \label{1}
f_{ijk}+v_{ikj}+g_{ijl}\beta _{lk}+u_{ikl}\alpha _{jl}=u_{jkl}\alpha
_{il}+v_{jki}
\end{equation}
\begin{equation}  \label{2}
f_{ijl}u_{lkm}+g_{ijl}l_{lkm}+u_{ikl}f_{jlm}+v_{ikl}u_{jlm}=u_{jkl}f_{ilm}+v_{jkl}u_{ilm}
\end{equation}
\begin{equation}  \label{3}
f_{ijl}v_{lkm}+g_{ijl}m_{lkm}+u_{ikl}g_{jlm}+v_{ikl}v_{jlm}=u_{jkl}g_{ilm}+v_{jkl}v_{ilm}
\end{equation}

Let us Introduce now the isotropy hypothesis. This requires that, provided $%
t_{ijk}$ is any among $f_{ijk}$, $g_{ijk}$, $l_{ijk}$, $m_{ijk}$, $u_{ijk}$
or $v_{ijk}$, the condition: $t_{ijk}=t\varepsilon _{ijk}$ be satisfied,
where $t$ is a scalar factor which is different ($f$, $g$, $l$, $m$, $u$ or $%
v$) in each case. But, that being true, (\ref{1})-(\ref{3}) are transformed
into:

\begin{equation}  \label{1a}
f\varepsilon _{ijk}+v\varepsilon _{ikj}+g\varepsilon _{ijl}\beta
_{lk}+u\varepsilon _{ikl}\alpha _{jl}=u\varepsilon _{jkl}\alpha
_{il}+v\varepsilon _{jki}
\end{equation}
\begin{equation}  \label{2a}
(fu+gl)\varepsilon _{ijl}\varepsilon _{lkm}+u(f+v)\varepsilon
_{ikl}\varepsilon _{jlm}=u(f+v)\varepsilon _{jkl}\varepsilon _{ilm}
\end{equation}
\begin{equation}  \label{3a}
(fv+gm)\varepsilon _{ijl}\varepsilon _{lkm}+(ug+v^2)\varepsilon
_{ikl}\varepsilon _{jlm}=(ug+v^2)\varepsilon _{jkl}\varepsilon _{ilm}
\end{equation}

that is:

\begin{equation}
f\varepsilon _{ijk}+g\varepsilon _{ijl}\beta _{lk}+u\varepsilon _{ikl}\alpha
_{jl}=u\varepsilon _{jkl}\alpha _{il}+2v\varepsilon _{ijk}  \label{1b}
\end{equation}
\[
(fu+gl)(\delta _{ik}\delta _{jm}-\delta _{im}\delta _{jk})+u(f+v)(\delta
_{im}\delta _{kj}-\delta _{ij}\delta _{km})=
\]
\begin{equation}
u(f+v)(\delta _{jm}\delta _{ki}-\delta _{ji}\delta _{km})  \label{2b}
\end{equation}

\[
(fv+gm)(\delta _{ik}\delta _{jm}-\delta _{im}\delta _{jk})+(ug+v^{2})(\delta
_{im}\delta _{kj}-\delta _{ij}\delta _{km})=
\]

\begin{equation}
(ug+v^{2})(\delta _{jm}\delta _{ki}-\delta _{ji}\delta _{km})  \label{3b}
\end{equation}

Let us consider now the subgroup of the Galilei group $SO(3)$, as being a
necessary invariance group for the commutation relations (\ref{c1})-(\ref{c3}%
). Such condition is satisfied by all canonical transformations of the form

\begin{equation}  \label{T1}
X_i^{\prime }=a_{ij}X_j
\end{equation}

\begin{equation}  \label{T2}
P_i^{\prime }=a_{ij}P_j
\end{equation}

with $a_{il}a_{jl}=\delta _{ij}$, leading us from (\ref{c1})-(\ref{c3}) to
the new ones

\[
\left[ X_{i}^{\prime },P_{j}^{\prime }\right] =
\]
\[
\left[ a_{il}X_{l},a_{jm}P_{m}\right] =a_{il}a_{jm}\left[ X_{l},P_{m}\right]
=a_{il}a_{jm}(i\delta _{lm}+iu\varepsilon _{lmk}X_{k}+iv\varepsilon
_{lmk}P_{k})=
\]
\[
ia_{il}a_{jl}+iua_{il}a_{jm}a_{nk}\varepsilon _{lmk}X_{n}^{\prime
}+iva_{il}a_{jm}a_{nk}\varepsilon _{lmk}P_{n}^{\prime }=
\]
\begin{equation}
i\delta _{ij}+iu\varepsilon _{ijn}X_{n}^{\prime }+iv\varepsilon
_{ijn}P_{n}^{\prime }  \label{A}
\end{equation}

\[
\left[ X_i^{\prime },X_j^{\prime }\right] = 
\]
\[
\left[ a_{il}X_l,a_{jm}X_m\right] =a_{il}a_{jm}\left[ X_l,X_m\right]
=a_{il}a_{jm}(i\alpha _{lm}+if\varepsilon _{lmk}X_k+ig\varepsilon
_{lmk}P_k)= 
\]

\[
ia_{il}a_{jm}\alpha _{lm}+ifa_{il}a_{jm}a_{nk}\varepsilon
_{lmk}X_{n}^{\prime }+iga_{il}a_{jm}a_{nk}\varepsilon _{lmk}P_{n}^{\prime }=
\]
\begin{equation}
ia_{il}a_{jm}\alpha _{lm}+if\varepsilon _{ijn}X_{n}^{\prime }+ig\varepsilon
_{ijn}P_{n}^{\prime }  \label{B}
\end{equation}

\[
\left[ P_{i}^{\prime },P_{j}^{\prime }\right] =
\]
\[
\left[ a_{il}P_{l},a_{jm}P_{m}\right] =a_{il}a_{jm}\left[ P_{l},P_{m}\right]
=a_{il}a_{jm}(i\beta _{lm}+il\varepsilon _{lmk}X_{k}+im\varepsilon
_{lmk}P_{k})=
\]
\[
ia_{il}a_{jm}\beta _{lm}+ila_{il}a_{jm}a_{nk}\varepsilon _{lmk}X_{n}^{\prime
}+ima_{il}a_{jm}a_{nk}\varepsilon _{lmk}P_{n}^{\prime }=
\]
\begin{equation}
ia_{il}a_{jm}\beta _{lm}+il\varepsilon _{ijn}X_{n}^{\prime }+im\varepsilon
_{ijn}P_{n}^{\prime }  \label{C}
\end{equation}

Now, we know the only isotropic form of a 2-index tensor with 3-valued
indexes is $\alpha _{ij}=0$ (resp. $\beta _{ij}=0$). Then

\begin{equation}  \label{cero(alfa-beta)}
\alpha _{ij}=\beta _{ij}=0
\end{equation}

In this way, equation (\ref{1b}) is reduced to

\begin{equation}  \label{1c}
f\varepsilon _{ijk}=2v\varepsilon _{ijk}
\end{equation}
from which

\begin{equation}  \label{(f-v)}
f=2v
\end{equation}

With these, equations (\ref{2b}) and (\ref{3b}) are somewhat simplified

\begin{equation}
(2uv+gl)(\delta _{ik}\delta _{jm}-\delta _{im}\delta _{jk})+3uv(\delta
_{im}\delta _{kj}-\delta _{ij}\delta _{km})=3uv(\delta _{jm}\delta
_{ki}-\delta _{ji}\delta _{km})  \label{2c}
\end{equation}
\[
(2v^{2}+gm)(\delta _{ik}\delta _{jm}-\delta _{im}\delta
_{jk})+(ug+v^{2})(\delta _{im}\delta _{kj}-\delta _{ij}\delta _{km})=
\]
\begin{equation}
(ug+v^{2})(\delta _{jm}\delta _{ki}-\delta _{ji}\delta _{km})  \label{3c}
\end{equation}

The argument so far exposed is completely analogous by permutating the
parameters, due to the fact that equations (\ref{c1})-(\ref{c3}) are
symmetrical with respect to the ``duality'' transformation

\begin{equation}  \label{Tdual}
\begin{array}{ccccc}
X_i\longrightarrow P_i, & u\longrightarrow v^{t(1,2)}=-v, & \alpha
\longrightarrow \beta , & f\longrightarrow m, & g\longrightarrow -l \\ 
P_i\longrightarrow -X_i, & v\longrightarrow u^{t(1,2)}=-u, & \beta
\longrightarrow \alpha , & m\longrightarrow -f, & l\longrightarrow g
\end{array}
\end{equation}
where $u^{t(1,2)}$ and $v^{t(1,2)}$ represent the three-index tensors $%
u_{ijk}$ and $v_{ijk}$ transposed with respect to their first two indices.
But, as we have already seen both tensors have to be proportional to the $%
\varepsilon $ tensor, all the equations are exactly the same as before with
the shuffling indicated in table (\ref{Tdual}). All this leads us to

\begin{equation}  \label{(m-u)}
m=-2u
\end{equation}
with which (\ref{3c}) is further simplified:

\[
(2v^{2}-2ug)(\delta _{ik}\delta _{jm}-\delta _{im}\delta
_{jk})+(ug+v^{2})(\delta _{im}\delta _{kj}-\delta _{ij}\delta _{km})=
\]
\begin{equation}
(ug+v^{2})(\delta _{jm}\delta _{ki}-\delta _{ji}\delta _{km})  \label{3e}
\end{equation}

By contracting in (\ref{2c}) and (\ref{3e}) $i$ with $k$ and $j$ with $m$: 
\begin{equation}  \label{r1}
gl=uv
\end{equation}
\begin{equation}  \label{r2}
v^2=3ug
\end{equation}

Other contractions produce either redundant conditions or identities (but no
contraction produces contra{\it di}ction!).

Let us try now with other forms of commutators:

Type $\left[ P,\left[ X,P\right] \right] $ (recall that already we have $%
\alpha =\beta =0$)$:$

\[
\left[ P_i,\left[ X_j,P_k\right] \right] =\left[ P_i,i\delta
_{jk}+iu\varepsilon _{jkl}X_l+iv\varepsilon _{jkl}P_l\right] =iu\varepsilon
_{jkl}\left[ P_i,X_l\right] +iv\varepsilon _{jkl}\left[ P_i,P_l\right] = 
\]

\[
-iu\varepsilon _{jkl}(i\delta _{li}+iu\varepsilon _{lik}X_k+iv\varepsilon
_{lik}P_k)+iv\varepsilon _{jkl}(il\varepsilon _{ilk}X_k+im\varepsilon
_{ilk}P_k)= 
\]

\[
u\varepsilon _{jki}+(u^2\varepsilon _{jkl}\varepsilon _{lik}-vl\varepsilon
_{jkl}\varepsilon _{ilk})X_k+(uv\varepsilon _{jkl}\varepsilon
_{lik}-vm\varepsilon _{jkl}\varepsilon _{ilk})P_k 
\]

On the other hand:

\[
\left[ P_i,\left[ X_j,P_k\right] \right] =\left[ P_i,X_jP_k\right] -\left[
P_i,P_kX_j\right] = 
\]

\[
\left[ P_i,X_j\right] P_k+X_j\left[ P_i,P_k\right] -\left[ P_i,P_k\right]
X_j-P_k\left[ P_i,X_j\right] = 
\]

\[
-(i\delta _{ji}+iu\varepsilon _{jil}X_l+iv\varepsilon
_{jil}P_l)P_k+X_j(il\varepsilon _{ikl}X_l+im\varepsilon _{ikl}P_l)- 
\]

\[
(il\varepsilon _{ikl}X_l+im\varepsilon _{ikl}P_l)X_j+P_k(i\delta
_{ji}+iu\varepsilon _{jil}X_l+iv\varepsilon _{jil}P_l)= 
\]

\[
iu\varepsilon _{jil}\left[ P_k,X_l\right] +iv\varepsilon _{jil}\left[ P_k,P_l%
\right] +il\varepsilon _{ikl}\left[ X_j,X_l\right] +im\varepsilon _{ikl}%
\left[ X_j,P_l\right] = 
\]

\[
u\varepsilon _{jik}+(u^2\varepsilon _{jil}\varepsilon _{lkm}-vl\varepsilon
_{jil}\varepsilon _{klm}-fl\varepsilon _{ikl}\varepsilon
_{jlm}-mu\varepsilon _{ikl}\varepsilon _{jlm})X_m+ 
\]

\[
(uv\varepsilon _{jil}\varepsilon _{lkm}-vm\varepsilon _{jil}\varepsilon
_{klm}-gl\varepsilon _{ikl}\varepsilon _{jlm}-mv\varepsilon
_{ikl}\varepsilon _{jlm})P_m 
\]

By equaling similar coefficients as before:

\[
u\varepsilon _{jik}=u\varepsilon _{jki} 
\]
from which the vanishing of $u$ is inferred. The argument is valid for $v$
as well, and then

\begin{equation}  \label{cero(u-v)}
u=v=0
\end{equation}
and, substituting in (\ref{r1})

\begin{equation}  \label{cero(g-l)}
gl=0
\end{equation}
from which one (or both) $g$ and $l$ has to vanish also.

But, as we have $u=v=0$, from (\ref{(m-u)}) and (\ref{(f-v)}), we have also

\begin{equation}  \label{cero(f-m)}
m=f=0
\end{equation}

But lets us make a brief recapitulation before closing the argument
motivating the present article. So far, the only commutation relations
compatible with spacial isotropy have been reduced to the form

\begin{equation}  \label{XP}
\left[ X_i,P_j\right] =i\delta _{ij}
\end{equation}

\begin{equation}  \label{XX}
\left[ X_i,X_j\right] =ig_{ijk}P_k
\end{equation}

\begin{equation}  \label{PP}
\left[ P_i,P_j\right] =il_{ijk}X_k
\end{equation}
where {\it one of both} constants, $g$ or $l$ is zero. The commutation
relations in the more general isotropy-compliant instance are then, either
on the one hand

\begin{equation}  \label{RC1}
\begin{array}{ccc}
\left[ X_i,P_j\right] =i\delta _{ij}, & \left[ X_i,X_j\right] =ig\varepsilon
_{ijk}P_k, & \left[ P_i,P_j\right] =0
\end{array}
\end{equation}
or else

\begin{equation}  \label{RC2}
\begin{array}{ccc}
\left[ X_i,P_j\right] =i\delta _{ij}, & \left[ X_i,X_j\right] =0, & \left[
P_i,P_j\right] =il\varepsilon _{ijk}X_k
\end{array}
\end{equation}

Both forms are invariant under the group of transformations (\ref{T1}), (\ref
{T2}). Both provide, besides, {\it a} {\it unique constant with dimensions
of length:} $L_1=g^{1/3}$ or else $L_2=l^{-1/3}$, thus both are attractive
as to the determination of a theory with good perspectives to deal with the
problem of gravitation.

Allow me to put forward the following observation as an argument that looks
quite upbeat in favour of the noncommutative-geometry idea before facing the
whole thing with a somewhat more critical mind. The question is to try to
decide oneself between both possibilities: the form given by (\ref{RC1}) is
particularly interesting due to a rather physical argument. The uncertainty
principle (\ref{incertidumbre2}) establishes for such election that

\begin{equation}  \label{incertidumbreXY}
\Delta _\psi X\Delta _\psi Y\geq \frac g2\left| \left\langle
P_z\right\rangle _\psi \right|
\end{equation}

This constraint has a beautiful and intuitive interpretation: if we prepare
a particle beam in a collision state such that

\begin{equation}  \label{ultravioleta}
\left\langle P_z\right\rangle _\psi >>\hbar g^{-1/3}
\end{equation}
then it will be in general impossible that it be arbitrarily small in the
transverse direction to the direction of collision, that is, with $\Delta
_\psi X\simeq \Delta _\psi Y\simeq 0$, due to the fact that, for (\ref
{incertidumbreXY}), we must have $\Delta _\psi X\Delta _\psi Y>>\hbar
g^{2/3}/2$. It is very reasonable to think of this relation as playing the
role of an effective cutoff for the ultraviolet range, as it precludes the
contribution of pointlike states in the transverse direction in the
dispersion relations. And it seems plausible to use this as a working
hypothesis for lack of a better phenomenological grasp of the full-fledged
noncommutative theory.

But let us look further beyond in the same direction we started from.
Indeed, we know we have not used all the information at our disposal to
maximally determine the structure constants. In order to resume, we apply
again the same procedure used to other third-order commutators, but now with
the commutation relations defining the algebra reduced to the form (\ref{RC1}%
). Let us assume, e.g., that we have $l=0$, $g\neq 0$. That is, we apply to
the ``promising so far'' commutation relations (\ref{RC1}) the iterated
commutation using Leibniz's rule in search of consistency (or
inconsistency!). On the one hand

\[
\left[ X_i,\left[ X_j,X_k\right] \right] =ig\varepsilon _{jkl}\left[ X_i,P_l%
\right] =ig\varepsilon _{jkl}(i\delta _{il})=-g\varepsilon _{jki} 
\]

But, on the other hand

\[
\left[ X_i,\left[ X_j,X_k\right] \right] =\left[ X_i,X_jX_k\right] -\left[
X_i,X_kX_j\right] = 
\]

\[
\left[ X_i,X_j\right] X_k+X_j\left[ X_i,X_k\right] -\left[ X_i,X_k\right]
X_j-X_k\left[ X_i,X_j\right] = 
\]

\[
(ig\varepsilon _{ijl}P_l)X_k+X_j(ig\varepsilon _{ikl}P_l)-(ig\varepsilon
_{ikl}P_l)X_j-X_k(ig\varepsilon _{ijl}P_l)=ig\varepsilon _{ijl}\left[ P_l,X_k%
\right] +ig\varepsilon _{ikl}\left[ X_j,P_l\right] = 
\]

\[
g\varepsilon _{ijk}-g\varepsilon _{ikj}=2g\varepsilon _{ijk} 
\]

By identifying, as before

\[
2g\varepsilon _{ijk}=-g\varepsilon _{jki}\Rightarrow g\varepsilon
_{ijk}=0\Rightarrow g=0 
\]

The argument is completely analogous to prove that $l=0$. Thus, the only
commutation relations of the general form (\ref{c1})-(\ref{c3}) consistent
with spacial isotropy in a first-quantised Galilean scenario, are
Heisenberg's; (\ref{heisenberg}).

\section{Final remarks}

With this work I have tried to put emphasis on the dangers involved when
dwelling in intrinsic formalisms without ever bringing the arguments down to
the level of particular coordinations. Valuable though such intrinsic
methods are, it is the particular coordinations that endorse any theory with
physical content. In this sense, it is easy to browse hundreds of pages
concerning noncommutative geometries without our eyes ever meeting a single
concrete expression of commutation relations in a definite, even just
nominally declared, coordinate system. It is sure that some arguments are 
{\it better} expressed in a particular coordinate system than in an
intrinsic manner or, at least, by starting out from such a coordinate system
in order to further consider the freedom the theory decrees by introducing a
transformation group.

On the other hand, this work is restricted to the study of the position and
momentum canonical operators on a configuration space of just one particle,
that is, it is focused to drawing a model of quantum mechanics on a flat
``noncommutative manifold''. In particular, it is not a study within the
context of quantum field theory nor does it contemplate the complications of
curvature and/or connection. It has to be interpreted, thus, as a mere
preliminary study. Today we know that any theory aspiring to some generality
must incorporate the feature of reducing itself to a field theory at low
energies. In consequence, a more serious attempt would imply the
generalisation of (\ref{c1})-(\ref{c3}) to the Fock space and refer them to
field operators. A possible attempt at generalising (\ref{RC1}) could be the
equal-time commutation relations (for a bosonic field)

\begin{equation}
\left[ \varphi _{a}({\bf x},0),\pi _{b}({\bf y},0)\right] =i\delta
_{ab}\delta ^{(3)}({\bf x}-{\bf y})  \label{F1}
\end{equation}
\begin{equation}
\left[ \varphi _{a}({\bf x},0),\varphi _{b}({\bf y},0)\right] =ig_{abc}\pi
_{c}({\bf x}-{\bf y})  \label{F2}
\end{equation}
\begin{equation}
\left[ \pi _{a}({\bf x},0),\pi _{b}({\bf y},0)\right] =0  \label{F3}
\end{equation}
that seems more promising than the ``naive'' version (\ref{RC1}) as the
isotropy argument cannot be exported to the inner space of field multiplets.
The problem now seems to be that (\ref{F1})-(\ref{F3}) have too many free
parameters with length dimensions for a general invariance group. The reader
can try these out and check that there is no easy way in which realtions (%
\ref{F1})-(\ref{F3}) can be be consistent and nonzero.

To end these comments, let as say that it is interesting (as an argumental
whetting of experimental appetite) to think that empirical checks on the
limits to the isotropy of space could serve as an indirect test for the
validity of the noncommuative model.

\end{document}